\documentclass[english]{article}
\usepackage[T1]{fontenc}
\usepackage[latin9]{inputenc}
\usepackage{amsbsy}
\usepackage{amssymb}
\usepackage{esint}

\makeatletter

\newcommand{\noun}[1]{\textsc{#1}}

\makeatother

\usepackage{babel}
\begin{document}

\title{Quantifying the Loss of Information from Binning List-Mode Data}

\author{Eric Clarkson}
\maketitle
\begin{abstract}
List-mode data is increasingly being uesd in SPECT and PET imaging,
among other imaging modalities. However, there are still many imaging
designs that effectively bin list-mode data before image reconstruction
or other estimation tasks are performed. Intuitively, the binning
operation should result in a loss of information. In this work we
show that this is true for Fisher information and provide a computational
method for quantifying the information loss. In the end we find that
the information loss depends on three factors. The first factor is
related to the smoothness of the mean data function for the list-mode
data. The second factor is the actual object being imaged. Finally,
the third factor is the binning scheme in relation to the other two
factors.
\end{abstract}

\section{Introduction}

Many imaging systems detect individual particles as they interact
with the imaging hardware. These particles are usually photons, but
there are also other choices such as neutrons, beta particles and
alpha particles. A list-mode imaging system produces an attribute
vector for each particle detected. The attribute vector may include
spatial position, energy, time or other attributes that can be assigned
to the particle {[}1-13{]}. When the particles are photons, list-mode
systems are also called photon processing systems to indicate that
the attributes are estimated from raw detector outputs via some data
processing algorithm {[}14-17{]}. In this work we are only concerned
with the fact that the imaging system produces an attribute vector
for each particle, regardless of how these attributes are arrived
at. 

We may envision the more common type of imaging system, a binned system,
as the result of resolving the space of all attribute vectors into
a collection of non-overlapping bins. The system then counts how many
attribute vectors fall into each bin and produces an integer vector
whose dimension is the number of bins. Intuitively, this would seem
to result in a loss of information. If we formulate the task of the
imaging system as the estimate of a certain number of parameters related
to the object being imaged, then we may consider quantifying this
loss of information, if indeed there is a loss of information.

If the parameter vector of interest has a known prior distribution,
then the Shannon information between the parameter vector and the
data may be used as a measure of information for the task at hand.
In this case the data processing inequality implies that the Shannon
information is not increased by the binning operation, but it does
not quantify the loss of information due to binning. In this work
we will use the Fisher Information Matrix (FIM) to quantify the information
loss due to binning. The FIM does not require a prior distribution
on the parameter vector of interest. We will show that the FIM always
decreases when list-mode data is binned and provide an expression
to calculate the information loss. We will find that the information
loss depends on three factors, the smoothness of the mean data function
for the list-mode data, the actual object being imaged, and the the
binning scheme in relation to the other two factors.

\section{List mode Fisher information}

We will confine our calculations to photon imaging systems where we
know that Poisson statistics are applicable. In a list mode imaging
system the data is a list of $q$-dimensional attribute vectors $\mathbf{a}_{n}$,
one for each photon detected. These photon attributes contained in
each of these vectors may include a two dimensional position on a
the face of a detector, the depth of interaction in a scintillation
detector, the energy of the photon, the direction the photon is travelling
when detected for a plenoptic array, and polarization parameters.
The collection of all possible attribute vectors is attribute space,
$\mathbb{A}$. We may arrange the data list into a matrix$\mathbf{A}=\left[\mathbf{a}_{1},\ldots\mathbf{a}_{N}\right]$
and, for a fixed exposure time, the conditional probability distribution
function (PDF) for the list is given by
\begin{equation}
pr\left(\mathbf{A}|\boldsymbol{\theta}\right)=\frac{\bar{N}\left(\boldsymbol{\theta}\right)}{N!}\exp\left[-\bar{N}\left(\boldsymbol{\theta}\right)\right]\left[\prod_{n=1}^{N}pr\left(\mathbf{a}_{n}|\boldsymbol{\theta}\right)\right],
\end{equation}
where $\boldsymbol{\theta}$ is a $p$-dimensional parameter vector
describing the object being imaged and $pr\left(\mathbf{a}|\boldsymbol{\theta}\right)$
is the attribute space conditional PDF determined by $\boldsymbol{\theta}$.
The specific form for $pr\left(\mathbf{a}|\boldsymbol{\theta}\right)$
depends on the imaging system. The FIM with respect to $\boldsymbol{\theta}$
for list mode data is defined by 
\begin{equation}
\mathbf{F}_{LM}\left(\boldsymbol{\theta}\right)=\left\langle \left[\nabla_{\boldsymbol{\theta}}\ln pr\left(\mathbf{A}|\boldsymbol{\theta}\right)\right]\left[\nabla_{\boldsymbol{\theta}}\ln pr\left(\mathbf{A}|\boldsymbol{\theta}\right)\right]^{\dagger}\right\rangle _{\mathbf{A}|\boldsymbol{\theta}}.
\end{equation}
Using the specific form for $pr\left(\mathbf{A}|\boldsymbol{\theta}\right)$
the list mode FIM can be written as 
\begin{equation}
\mathbf{F}_{LM}\left(\boldsymbol{\theta}\right)=\bar{N}\left(\boldsymbol{\theta}\right)\left\{ \left\langle \left[\nabla_{\boldsymbol{\theta}}\ln pr\left(\mathbf{a}|\boldsymbol{\theta}\right)\right]\left[\nabla_{\boldsymbol{\theta}}\ln pr\left(\mathbf{a}|\boldsymbol{\theta}\right)\right]^{\dagger}\right\rangle _{\mathbf{a}|\boldsymbol{\theta}}+\left[\nabla_{\boldsymbol{\theta}}\ln\bar{N}\left(\boldsymbol{\theta}\right)\right]\left[\nabla_{\boldsymbol{\theta}}\ln\bar{N}\left(\boldsymbol{\theta}\right)\right]^{\dagger}\right\} .
\end{equation}
This is a $p\times p$ matrix which figures prominently in the task
of estimating $\boldsymbol{\theta}$ from the data list $\mathbf{A}$
via the Cramer-Rao bound {[}18{]}. As we will discuss further below,
the FIM is also related to the performance of an ideal observer using
the data list $\mathbf{A}$ for the task of detecting a change in
the parameter vector from $\boldsymbol{\theta}$ to $\boldsymbol{\theta}+\triangle\boldsymbol{\theta}$
{[}19,20{]}.

\section{Binned Fisher information}

List mode data can also de described as a Poisson Point Process {[}21{]}
on attribute space via the generalized function $g\left(\mathbf{a}\right)$
given by
\begin{equation}
g\left(\mathbf{a}\right)=\sum_{n=1}^{N}\delta\left(\mathbf{a}-\mathbf{a}_{n}\right).
\end{equation}
If we introduce binning functions $b_{m}\left(\mathbf{a}\right)$
for $m=1,\ldots,M$, then we get the components 
\begin{equation}
g_{m}=\sum_{n=1}^{N}b_{m}\left(\mathbf{a}_{n}\right)=\int_{\mathbb{A}}g\left(\mathbf{a}\right)b_{m}\left(\mathbf{a}\right)d^{q}a
\end{equation}
of a binned $M$-dimensional data vector $\mathbf{g}$. We will assume
that the functions $b_{m}\left(\mathbf{a}\right)$ are binary with
non-overlapping supports, so that $b_{m}\left(\mathbf{a}\right)b_{m'}\left(\mathbf{a}\right)=\delta_{mm'}b_{m}\left(\mathbf{a}\right)$,
and that they cover all of attribute space, i.e. for all $\mathbf{a}$in
$\mathbb{A}$ we have $b_{1}\left(\mathbf{a}\right)+\ldots+b_{M}\left(\mathbf{a}\right)=1.$
The PDF for the binned data vector is multivariate Poisson:
\begin{equation}
pr\left(\mathbf{g}|\boldsymbol{\theta}\right)=\prod_{m=1}^{M}\frac{\left[\bar{g}_{m}\left(\boldsymbol{\theta}\right)\right]^{g_{m}}}{g_{m}!}\exp\left[-\bar{g}_{m}\left(\boldsymbol{\theta}\right)\right]
\end{equation}
with $\bar{g}_{m}\left(\boldsymbol{\theta}\right)=\left\langle g_{m}\right\rangle _{\mathbf{A}|\boldsymbol{\theta}}.$
The binned FIM is defined by
\begin{equation}
\mathbf{F}_{B}\left(\boldsymbol{\theta}\right)=\left\langle \left[\nabla_{\theta}\ln pr\left(\mathbf{g}|\boldsymbol{\theta}\right)\right]\left[\nabla_{\theta}\ln pr\left(\mathbf{g}|\boldsymbol{\theta}\right)\right]^{\dagger}\right\rangle 
\end{equation}
This matrix is relevant to the task of estimating $\boldsymbol{\theta}$
from the data vector $\mathbf{g}$ via the corresponding Cramer-Rao
bound. As above, this FIM is also related to the performance of an
ideal observer using the data vector $\mathbf{g}$ for the task of
detecting a change in the parameter vector from $\boldsymbol{\theta}$
to $\boldsymbol{\theta}+\triangle\boldsymbol{\theta}$. Intuitively
we expect better perfomance on the estimation task or the detection
task with the list mode data than with the binned data, since there
is an obvious loss of information about each photon in the transition
from list mode to binned data. In the following we will show that
this is true and derive an equation that quantifies this loss of information
using the FIM matrices for the two data types.

\section{Relation between the two FIMs}

The attribute-space PDF can be written in terms of the conditional
mean of the Poisson Point Process $\bar{g}\left(\mathbf{a}|\boldsymbol{\theta}\right)=\left\langle g\left(\mathbf{a}\right)\right\rangle _{\mathbf{A}|\boldsymbol{\theta}}$
via the equation $\bar{N}\left(\boldsymbol{\theta}\right)pr\left(\mathbf{a}|\boldsymbol{\theta}\right)=\bar{g}\left(\mathbf{a}|\boldsymbol{\theta}\right)$,
where 
\begin{equation}
\bar{N}\left(\boldsymbol{\theta}\right)=\int_{\mathbb{A}}\bar{g}\left(\mathbf{a}|\boldsymbol{\theta}\right)d^{q}a=\sum_{m=1}^{M}\bar{g}_{m}\left(\boldsymbol{\theta}\right).
\end{equation}
Now we may write the list mode FIM as
\begin{equation}
\mathbf{F}_{LM}\left(\boldsymbol{\theta}\right)=\bar{N}\left(\boldsymbol{\theta}\right)\left\langle \left[\nabla_{\boldsymbol{\theta}}\ln\bar{g}\left(\mathbf{a}|\boldsymbol{\theta}\right)\right]\left[\nabla_{\boldsymbol{\theta}}\ln\bar{g}\left(\mathbf{a}|\boldsymbol{\theta}\right)\right]^{\dagger}\right\rangle _{\mathbf{a}|\boldsymbol{\theta}},
\end{equation}
and this is the same as the integral expression
\begin{equation}
\mathbf{F}_{LM}\left(\boldsymbol{\theta}\right)=\int_{\mathbb{A}}\left[\nabla_{\boldsymbol{\theta}}\ln\bar{g}\left(\mathbf{a}|\boldsymbol{\theta}\right)\right]\left[\nabla_{\boldsymbol{\theta}}\ln\bar{g}\left(\mathbf{a}|\boldsymbol{\theta}\right)\right]^{\dagger}\bar{g}\left(\mathbf{a}|\boldsymbol{\theta}\right)d^{q}a.
\end{equation}
Thus the list mode FIM is determined entirely by the conditional mean
function $\bar{g}\left(\mathbf{a}|\boldsymbol{\theta}\right)$. 

Meanwhile, we have the relation between conditional means
\begin{equation}
\bar{g}_{m}\left(\boldsymbol{\theta}\right)=\int_{\mathbb{A}}b_{m}\left(\mathbf{a}\right)\bar{g}\left(\mathbf{a}|\boldsymbol{\theta}\right)d^{q}a
\end{equation}
and we can define a finite conditional probability distribution $Pr\left(m|\boldsymbol{\theta}\right)$
on $\left\{ 1,\ldots,M\right\} $ via $\bar{N}\left(\boldsymbol{\theta}\right)Pr\left(m|\boldsymbol{\theta}\right)=\bar{g}_{m}\left(\boldsymbol{\theta}\right)$.
Now the binned FIM is given by an expectation with respect to this
finite probability distribution
\begin{equation}
\mathbf{F}_{B}\left(\boldsymbol{\theta}\right)=\bar{N}\left(\boldsymbol{\theta}\right)\left\langle \left[\nabla_{\boldsymbol{\theta}}\ln\bar{g}_{m}\left(\boldsymbol{\theta}\right)\right]\left[\nabla_{\boldsymbol{\theta}}\ln\bar{g}_{m}\left(\boldsymbol{\theta}\right)\right]^{\dagger}\right\rangle _{m|\boldsymbol{\theta}}.
\end{equation}
Notice the similarity with the corresponding expectation expression
for the list mode FIM. The only difference is that a PDF for the attribute
vector $\mathbf{a}$ has been replace by a finite probability distribution
for the bin index $m.$ The binned FIM can also be written as
\begin{equation}
\mathbf{F}_{B}\left(\boldsymbol{\theta}\right)=\sum_{m=1}^{M}\left[\nabla_{\boldsymbol{\theta}}\ln\bar{g}_{m}\left(\boldsymbol{\theta}\right)\right]\left[\nabla_{\boldsymbol{\theta}}\ln\bar{g}_{m}\left(\boldsymbol{\theta}\right)\right]^{\dagger}\bar{g}_{m}\left(\boldsymbol{\theta}\right)
\end{equation}
Thus, from one viewpoint , we get the binned FIM by using the bin
functions $b_{m}\left(\mathbf{a}\right)$ to numerically perform the
integration in the list mode FIM. It is not obvious at this point
that this numerical procedure will always produce a lower value for
the FIM. 

The ideal observer detectability $d\left(\boldsymbol{\theta}_{0},\boldsymbol{\theta}_{1}\right)$
for the task of detecting a change in the parameter vector from $\boldsymbol{\theta}_{0}$
to $\boldsymbol{\theta}_{1}$ is defined by 
\begin{equation}
AUC\left(\boldsymbol{\theta}_{0},\boldsymbol{\theta}_{1}\right)=\frac{1}{2}+\frac{1}{2}\mathrm{erf}\left[\frac{1}{2}d\left(\boldsymbol{\theta}_{0},\boldsymbol{\theta}_{1}\right)\right],
\end{equation}
where $AUC\left(\boldsymbol{\theta}_{0},\boldsymbol{\theta}_{1}\right)$
is the area under the ROC curve for the ideal observer. It has been
shown that, to lowest order, $d^{2}\left(\boldsymbol{\theta},\boldsymbol{\theta}+\triangle\boldsymbol{\theta}\right)=\triangle\boldsymbol{\theta}^{\dagger}\mathbf{F}\left(\boldsymbol{\theta}\right)\triangle\boldsymbol{\theta}+\ldots,$where
$\mathbf{F}\left(\boldsymbol{\theta}\right)$ is the FIM for the conditional
PDF of the data. Thus the scalar 
\begin{equation}
\triangle\boldsymbol{\theta}^{\dagger}\mathbf{F}_{LM}\left(\boldsymbol{\theta}\right)\triangle\boldsymbol{\theta}=\int_{\mathbb{A}}\left[\triangle\boldsymbol{\theta}^{\dagger}\nabla_{\boldsymbol{\theta}}\bar{g}\left(\mathbf{a}|\boldsymbol{\theta}\right)\right]^{2}\left[\bar{g}\left(\mathbf{a}|\boldsymbol{\theta}\right)\right]^{-1}d^{q}a
\end{equation}
gives the square of the approximate ideal-observer detectability for
this task when we use list mode data. Similarly, the scalar
\begin{equation}
\triangle\boldsymbol{\theta}^{\dagger}\mathbf{F}_{B}\left(\boldsymbol{\theta}\right)\triangle\boldsymbol{\theta}=\sum_{m=1}^{M}\left[\int_{\mathbb{A}}b_{m}\left(\mathbf{a}\right)\triangle\boldsymbol{\theta}^{\dagger}\nabla_{\boldsymbol{\theta}}\bar{g}\left(\mathbf{a}|\boldsymbol{\theta}\right)d^{q}a\right]^{2}\left[\bar{g}_{m}\left(\boldsymbol{\theta}\right)\right]^{-1}
\end{equation}
gives the square of the approximate ideal-observer detectability for
a small change in the parameter vector from $\boldsymbol{\theta}$
to $\boldsymbol{\theta}+\triangle\boldsymbol{\theta}$ if we are using
binned data. We will show that $\triangle\boldsymbol{\theta}^{\dagger}\mathbf{F}_{LM}\left(\boldsymbol{\theta}\right)\triangle\boldsymbol{\theta}\geq\triangle\boldsymbol{\theta}^{\dagger}\mathbf{F}_{B}\left(\boldsymbol{\theta}\right)\triangle\boldsymbol{\theta}$
for all $\boldsymbol{\theta}$ and $\triangle\boldsymbol{\theta}$.
By definition, this then implies that $\mathbf{F}_{LM}\left(\boldsymbol{\theta}\right)\geq\mathbf{F}_{B}\left(\boldsymbol{\theta}\right)$
as matrices for all $\boldsymbol{\theta}.$

To simplify the calculations we will define $\gamma\left(\mathbf{a}\right)=\triangle\boldsymbol{\theta}^{\dagger}\nabla_{\boldsymbol{\theta}}\bar{g}\left(\mathbf{a}|\boldsymbol{\theta}\right)$
and suppress the fact that this function also depends on $\boldsymbol{\theta}$
and $\triangle\boldsymbol{\theta}$, since these vectors are fixed
for the purposes of this computation. Then we have
\begin{equation}
\triangle\boldsymbol{\theta}^{\dagger}\mathbf{F}_{LM}\left(\boldsymbol{\theta}\right)\triangle\boldsymbol{\theta}=\int_{\mathbb{A}}\left[\gamma\left(\mathbf{a}\right)\right]^{2}\left[\bar{g}\left(\mathbf{a}|\boldsymbol{\theta}\right)\right]^{-1}d^{q}a
\end{equation}
This expression suggests that, for fixed $\boldsymbol{\theta}$, we
define a weighted Hilbert space inner product for functions on attribute
space via
\begin{equation}
\left(\gamma,\gamma'\right)_{\boldsymbol{\theta}}=\int_{\mathbb{A}}\gamma^{\ast}\left(\mathbf{a}\right)\gamma'\left(\mathbf{a}\right)\left[\bar{g}\left(\mathbf{a}|\boldsymbol{\theta}\right)\right]^{-1}d^{q}a=\left(\gamma,\mathcal{D}_{\boldsymbol{\theta}}^{-1}\gamma'\right)
\end{equation}
where $\mathcal{D}_{\boldsymbol{\theta}}^{-1}\gamma'\left(\mathbf{a}\right)=\gamma'\left(\mathbf{a}\right)\left[\bar{g}\left(\mathbf{a}|\boldsymbol{\theta}\right)\right]^{-1}$.
The list-mode approximate detectability is then given by the corresponding
weighted Hilbert-space norm as $\triangle\boldsymbol{\theta}^{\dagger}\mathbf{F}_{LM}\left(\boldsymbol{\theta}\right)\triangle\boldsymbol{\theta}=\left\Vert \gamma\right\Vert _{\boldsymbol{\theta}}^{2}$.

For the binned data we have the summation
\begin{equation}
\triangle\boldsymbol{\theta}^{\dagger}\mathbf{F}_{B}\left(\boldsymbol{\theta}\right)\triangle\boldsymbol{\theta}=\sum_{m=1}^{M}\left[\int_{\mathbb{A}}b_{m}\left(\mathbf{a}\right)\gamma\left(\mathbf{a}\right)d^{q}a\right]^{2}\left[\bar{g}_{m}\left(\boldsymbol{\theta}\right)\right]^{-1}.
\end{equation}
We define the binning operator $\mathcal{B}$ by
\begin{equation}
\left(\mathcal{B}\gamma\right)_{m}=\int_{\mathbb{A}}b_{m}\left(\mathbf{a}\right)\gamma\left(\mathbf{a}\right)d^{q}a
\end{equation}
and the ordinary Hilbert space adjoint of this operator by
\begin{equation}
\mathcal{B^{\dagger}}\mathbf{g}=\sum_{m=1}^{M}g_{m}b_{m}\left(\mathbf{a}\right).
\end{equation}
Then we have a simpler looking expression
\begin{equation}
\triangle\boldsymbol{\theta}^{\dagger}\mathbf{F}_{B}\left(\boldsymbol{\theta}\right)\triangle\boldsymbol{\theta}=\sum_{m=1}^{M}\left(\mathcal{B}\gamma\right)_{m}^{2}\left[\bar{g}_{m}\left(\boldsymbol{\theta}\right)\right]^{-1}.
\end{equation}
This expression suggests introducing a weighted inner product in the
$M$-dimensional data space by
\begin{equation}
\left(\mathbf{g},\mathbf{g}'\right)_{\boldsymbol{\theta}}=\sum_{m=1}^{M}g_{m}^{*}g'_{m}\left[\bar{g}_{m}\left(\boldsymbol{\theta}\right)\right]^{-1}=\left(\mathbf{g},\mathbf{D}_{\boldsymbol{\theta}}^{-1}\mathbf{g}'\right)
\end{equation}
where $\mathbf{D}_{\boldsymbol{\theta}}^{-1}$ is a diagonal $M\times M$
matrix with the numbers $\left[\bar{g}_{m}\left(\boldsymbol{\theta}\right)\right]^{-1}$
along the diagonal. With this notation the binned approxiamte detectability
is given by the weighted norm $\triangle\boldsymbol{\theta}^{\dagger}\mathbf{F}_{B}\left(\boldsymbol{\theta}\right)\triangle\boldsymbol{\theta}=\left\Vert \mathcal{B}\gamma\right\Vert _{\boldsymbol{\theta}}^{2}$.
Thus both $\triangle\boldsymbol{\theta}^{\dagger}\mathbf{F}_{LM}\left(\boldsymbol{\theta}\right)\triangle\boldsymbol{\theta}$
and $\triangle\boldsymbol{\theta}^{\dagger}\mathbf{F}_{B}\left(\boldsymbol{\theta}\right)\triangle\boldsymbol{\theta}$
are now expressed as weighted Hilbert space norms of the function
$\gamma$ and the vector $\mathcal{B}\gamma$, respectively. 

We can now think of the binning operator as a map between two weighted
Hilbert spaces: $\mathcal{B}:L_{\boldsymbol{\theta}}^{2}\left(\mathbb{A}\right)\longrightarrow\mathbb{R}_{\boldsymbol{\theta}}^{M}$.
As a first step we want to find the pseudoinverse of this operator.
We begin by finding the adjoint of this operator. Note that this is
not the ``ordinary adjoint'' described above. The relevant calculation
for this adjoint is given by
\begin{equation}
\left(\mathbf{g},\mathcal{B}\gamma'\right)_{\boldsymbol{\theta}}=\left(\gamma,\mathbf{D}_{\boldsymbol{\theta}}^{-1}\mathcal{B}\gamma'\right)=\left(\mathcal{B}^{\dagger}\mathbf{D}_{\boldsymbol{\theta}}^{-1}\mathbf{g},\gamma'\right)=\left(\mathcal{D}_{\boldsymbol{\theta}}\mathcal{B}^{\dagger}\mathbf{D}_{\boldsymbol{\theta}}^{-1}\mathbf{g},\mathcal{D}_{\boldsymbol{\theta}}^{-1}\gamma'\right)=\left(\mathcal{D}_{\boldsymbol{\theta}}\mathcal{B}^{\dagger}\mathbf{D}_{\boldsymbol{\theta}}^{-1}\mathbf{g},\gamma'\right)_{\boldsymbol{\theta}}
\end{equation}
Thus $\mathcal{D}_{\boldsymbol{\theta}}\mathcal{B}^{\dagger}\mathbf{D}_{\boldsymbol{\theta}}^{-1}$
is the adjoint operator we are looking for. The pseudoinverse of $\mathcal{B}$,
as an operator between the weighted Hilbert spaces, is then given
by
\begin{equation}
\mathcal{B}^{+}=\mathcal{D}_{\boldsymbol{\theta}}\mathcal{B^{\dagger}}\mathbf{D}_{\boldsymbol{\theta}}^{-1}\left(\mathcal{B}\mathcal{D}_{\boldsymbol{\theta}}\mathcal{B^{\dagger}}\mathbf{D}_{\boldsymbol{\theta}}^{-1}\right)^{-1}=\mathcal{D}_{\boldsymbol{\theta}}\mathcal{B^{\dagger}}\left(\mathcal{B}\mathcal{D}_{\boldsymbol{\theta}}\mathcal{B^{\dagger}}\right)^{-1}
\end{equation}
If we look at this expression in detail we first note that
\begin{equation}
\mathcal{D}_{\boldsymbol{\theta}}\mathcal{B^{\dagger}}\mathbf{g}\left(\mathbf{a}\right)=\mathcal{D}_{\boldsymbol{\theta}}\left\{ \sum_{m'=1}^{M}g_{m'}b_{m'}\left(\mathbf{a}\right)\right\} =\bar{g}\left(\mathbf{a}|\boldsymbol{\theta}\right)\sum_{m'=1}^{M}g_{m'}b_{m'}\left(\mathbf{a}\right)
\end{equation}
Now implementing the binning operator, and using the properties of
the binning functions, gives us, in component form,
\begin{equation}
\left(\mathcal{B}\mathcal{D}_{f}\mathcal{B^{\dagger}}\mathbf{g}\right)_{m}=g_{m}\int_{\mathbb{A}}\bar{g}\left(\mathbf{a}|\boldsymbol{\theta}\right)b_{m}\left(\mathbf{a}\right)d^{q}a=g_{m}\bar{g}_{m}\left(\boldsymbol{\theta}\right).
\end{equation}
Therefore we find that $\mathcal{B}\mathcal{D}_{\boldsymbol{\theta}}\mathcal{B^{\dagger}}=\mathbf{D}_{\boldsymbol{\theta}}$.
Now we have a simplified version of the needed pseudoinverse: $\mathcal{B}^{+}=\mathcal{D}_{\boldsymbol{\theta}}\mathcal{B^{\dagger}}\mathbf{D}_{\boldsymbol{\theta}}^{-1}$. 

We may decompose the function $\gamma$ into two components $\gamma=\gamma_{1}+\gamma_{0}$,
where $\gamma_{0}$ is a null function with respect to the binning
operator, i.e. $\mathcal{B}\gamma_{0}=\mathbf{0}$, and we have the
orthogonality condition $\left(\gamma_{1},\gamma_{0}\right)_{\boldsymbol{\theta}}=0.$
The component $\gamma_{1}$ is given by$\gamma_{1}=\mathcal{B}^{+}\mathcal{B}\gamma$.
Therefore we have $\gamma_{1}=\mathcal{D}_{\boldsymbol{\theta}}\mathcal{B^{\dagger}}\mathbf{D}_{\boldsymbol{\theta}}^{-1}\mathcal{B}\gamma$.
Writing this equation out in detail we have
\begin{equation}
\gamma_{1}\left(\mathbf{a}\right)=\mathcal{B}^{+}\mathcal{B}\gamma\left(\mathbf{a}\right)=\bar{g}\left(\mathbf{a}|\boldsymbol{\theta}\right)\sum_{m=1}^{M}\left[\frac{\left(\mathcal{B}\gamma\right)_{m}}{\bar{g}_{m}\left(\boldsymbol{\theta}\right)}\right]b_{m}\left(\boldsymbol{a}\right).
\end{equation}
The null component of $\gamma$ is then defined by $\gamma_{0}\left(\mathbf{a}\right)=\gamma\left(\mathbf{a}\right)-\gamma_{1}\left(\mathbf{a}\right)$,
and due to the orthogonality condition we have $\left\Vert \gamma\right\Vert _{\boldsymbol{\theta}}^{2}=\left\Vert \gamma_{1}\right\Vert _{\boldsymbol{\theta}}^{2}+\left\Vert \gamma_{0}\right\Vert _{\boldsymbol{\theta}}^{2}$.

Now we examine the square magnitude, in the weighted Hilbert space,
of the $\gamma_{1}$ component of $\gamma$:
\begin{equation}
\left\Vert \gamma_{1}\right\Vert _{\boldsymbol{\theta}}^{2}=\int_{\mathbb{A}}\left[\gamma_{1}\left(\mathbf{a}\right)\right]^{2}\left[\bar{g}\left(\mathbf{a}|\boldsymbol{\theta}\right)\right]^{-1}d^{q}a.
\end{equation}
Substituting our expression for $\gamma_{1}\left(\mathbf{a}\right)$
and then using the properties of the binning functions, we find that
\begin{equation}
\left\Vert \gamma_{1}\right\Vert _{\boldsymbol{\theta}}^{2}=\sum_{m=1}^{M}\left[\frac{\left(\mathcal{B}\gamma\right)_{m}}{\bar{g}_{m}\left(\boldsymbol{\theta}\right)}\right]^{2}\int_{\mathbb{A}}\bar{g}\left(\mathbf{a}|\boldsymbol{\theta}\right)b_{m}\left(\boldsymbol{a}\right)d^{q}a.
\end{equation}
After performing the integration we find that $\left\Vert \gamma_{1}\right\Vert _{\boldsymbol{\theta}}^{2}=\triangle\boldsymbol{\theta}^{\dagger}\mathbf{F}_{B}\left(\boldsymbol{\theta}\right)\triangle\boldsymbol{\theta}$. 

Now we can find the difference between the list-mode and binned approximate
detectabilities
\begin{equation}
\triangle\boldsymbol{\theta}^{\dagger}\mathbf{F}_{LM}\left(\boldsymbol{\theta}\right)\triangle\boldsymbol{\theta}-\triangle\boldsymbol{\theta}^{\dagger}\mathbf{F}_{B}\left(\boldsymbol{\theta}\right)\triangle\boldsymbol{\theta}=\left\Vert \gamma_{0}\right\Vert _{\boldsymbol{\theta}}^{2}.
\end{equation}
Using the definition of $\gamma\left(\mathbf{a}\right)$ we have the
final result
\begin{equation}
\triangle\boldsymbol{\theta}^{\dagger}\mathbf{F}_{LM}\left(\boldsymbol{\theta}\right)\triangle\boldsymbol{\theta}-\triangle\boldsymbol{\theta}^{\dagger}\mathbf{F}_{B}\left(\boldsymbol{\theta}\right)\triangle\boldsymbol{\theta}=\int_{\mathbb{A}}\left\{ \left[\triangle\boldsymbol{\theta}^{\dagger}\nabla_{\boldsymbol{\theta}}\bar{g}\left(\mathbf{a}|\boldsymbol{\theta}\right)\right]_{0}\right\} ^{2}\left[\bar{g}\left(\mathbf{a}|\boldsymbol{\theta}\right)\right]^{-1}d^{q}a.
\end{equation}
Since $\triangle\boldsymbol{\theta}$ is arbitrary, this equation
gives us a matrix inequality between FIMs :$\mathbf{F}_{LM}\left(\boldsymbol{\theta}\right)\geq\mathbf{F}_{B}\left(\boldsymbol{\theta}\right)$
with equality only if $\left[\triangle\boldsymbol{\theta}^{\dagger}\nabla_{\boldsymbol{\theta}}\bar{g}\left(\mathbf{a}|\boldsymbol{\theta}\right)\right]_{0}=0.$
The equality condition can also be written as
\begin{equation}
\gamma\left(\mathbf{a}\right)=\bar{g}\left(\mathbf{a}|\boldsymbol{\theta}\right)\sum_{m=1}^{M}\left[\frac{\left(\mathcal{B}\gamma\right)_{m}}{\bar{g}_{m}\left(\boldsymbol{\theta}\right)}\right]b_{m}\left(\boldsymbol{a}\right)
\end{equation}
where $\gamma\left(\mathbf{a}\right)=\triangle\boldsymbol{\theta}^{\dagger}\nabla_{\boldsymbol{\theta}}\bar{g}\left(\mathbf{a}|\boldsymbol{\theta}\right)$
and 
\begin{equation}
\left(\mathcal{B}\gamma\right)_{m}=\int_{\mathbb{A}}b_{m}\left(\mathbf{a}\right)\triangle\boldsymbol{\theta}^{\dagger}\nabla_{\boldsymbol{\theta}}\bar{g}\left(\mathbf{a}|\boldsymbol{\theta}\right)d^{q}a.
\end{equation}
The probability is zero that this condition will be satisfied in any
real imaging situation, which means that binning always results in
a loss of Fisher information.

Note that the condition for no loss of Fisher information due to binning
can be written as 
\begin{equation}
\gamma_{0}\left(\mathbf{a}\right)=\bar{g}\left(\mathbf{a}|\boldsymbol{\theta}\right)\sum_{m=1}^{M}\left[\frac{\gamma\left(\mathbf{a}\right)}{\bar{g}\left(\mathbf{a}|\boldsymbol{\theta}\right)}-\frac{\left(\mathcal{B}\gamma\right)_{m}}{\bar{g}_{m}\left(\boldsymbol{\theta}\right)}\right]b_{m}\left(\boldsymbol{a}\right)=0.
\end{equation}
This then gives us
\begin{equation}
\triangle\boldsymbol{\theta}^{\dagger}\mathbf{F}_{LM}\left(\boldsymbol{\theta}\right)\triangle\boldsymbol{\theta}-\triangle\boldsymbol{\theta}^{\dagger}\mathbf{F}_{B}\left(\boldsymbol{\theta}\right)\triangle\boldsymbol{\theta}=\sum_{m=1}^{M}\int_{\mathbb{A}}\left[\frac{\gamma\left(\mathbf{a}\right)}{\bar{g}\left(\mathbf{a}|\boldsymbol{\theta}\right)}-\frac{\left(\mathcal{B}\gamma\right)_{m}}{\bar{g}_{m}\left(\boldsymbol{\theta}\right)}\right]^{2}\bar{g}\left(\mathbf{a}|\boldsymbol{\theta}\right)b_{m}\left(\boldsymbol{a}\right)d^{q}a.
\end{equation}
Thus each bin contributes an amount to the loss of detectability according
to three factors. The first factor is the deviation of the quantity
in curly brackets from zero within that bin. The second factor is
the value of the mean data function $\bar{g}\left(\mathbf{a}|\boldsymbol{\theta}\right)$
within the bin. The third factor is the size of the bin itself. Therefore
the efficiency of any particular choice of bins in preserving Fisher
information depends on the actual parameter value $\boldsymbol{\theta}$
as well as the bin sizes. Having derived this relationship it is actually
streightforward to prove that it is valid without any discussion of
weighted Hilbert spaces. However, the path we followed to get here
demonstrates that the loss of Fisher information due to binning is
caused by the null space of the binning operator $\mathcal{B}:L_{\boldsymbol{\theta}}^{2}\left(\mathbb{A}\right)\longrightarrow\mathbb{R}_{\boldsymbol{\theta}}^{M}$,
when viewed as an operator between weighted Hilbert spaces.

\section{FIMs for object Reconstruction}

In this section the parameter vector $\boldsymbol{\theta}$ is replaced
with a function $f\left(\mathbf{r}\right)$ of spatial coordinates.
This complication is mitigated by a linear relation between the object
function and mean data function via a linear operator:
\begin{equation}
\bar{g}\left(\mathbf{a}|f\right)=\mathcal{L}f\left(\mathbf{a}\right)=\int_{S}L\left(\mathbf{a},\mathbf{r}\right)f\left(\mathbf{r}\right)d^{q}r,
\end{equation}
where $S$ is a support region for object functions in a $q$-dimensional
space. The gradient operator $\nabla_{\boldsymbol{\theta}}$ is replaced
by a functional derivative or Frechet derivative. The FIM matrices
are now a Fisher information operators $\mathcal{F}_{LM}$ and $\mathcal{F}_{B}$.
The simplicity of the connection between $f\left(\mathbf{r}\right)$
and $\bar{g}\left(\mathbf{a}|f\right)$ makes the functional derivative
easy to compute. 

The end result for the detectability calculation with list-mode data
is then given by
\begin{equation}
\left(\triangle f,\mathcal{F}_{LM}\left(f\right)\triangle f\right)=\int_{\mathbb{A}}\left[\mathcal{L}\triangle f\left(\mathbf{a}\right)\right]^{2}\left[\mathcal{L}f\left(\mathbf{a}\right)\right]^{-1}d^{q}a.
\end{equation}
The weighted inner product for functions on attribute space is now
defined by
\begin{equation}
\left(g,g'\right)_{f}=\left(g,\mathcal{D}_{f}^{-1}g'\right)=\int_{\mathbb{A}}g^{*}\left(\mathbf{a}\right)g'\left(\mathbf{a}\right)\left[\mathcal{L}f\left(\mathbf{a}\right)\right]^{-1}d^{q}a
\end{equation}
With the resulting weighted Hilbert space norm we then have $\left(\triangle f,\mathcal{F}_{LM}\left(f\right)\triangle f\right)=\left\Vert \mathcal{L}\triangle f\right\Vert _{f}^{2}.$

The imaging operator for the binned imaging system is given by the
concatenation of the list-mode system operator with the binning operator:$\mathcal{H}=\mathcal{B}\mathcal{L}$.
The detectability calculation for the binned system then gives us
\begin{equation}
\left(\triangle f,\mathcal{F}_{B}\left(f\right)\triangle f\right)=\sum_{m=1}^{M}\left[\left(\mathcal{H}\triangle f\right)\right]_{m}^{2}\left[\mathcal{H}f\right]_{m}^{-1}.
\end{equation}
As before we introduce a weighted inner product in data space via
\begin{equation}
\left(\mathbf{g},\mathbf{g}'\right)_{f}=\left(\mathbf{g},\mathbf{D}_{f}^{-1}\mathbf{g}'\right)=\sum_{m=1}^{M}g_{m}^{*}g'_{m}\left[\mathcal{H}f\right]_{m}^{-1},
\end{equation}
and we then have $\left(\triangle f,\mathcal{F}_{B}\left(f\right)\triangle f\right)=\left\Vert \mathcal{H}\triangle f\right\Vert _{f}^{2}$.

The relevant operators are now the list mode system operator $\mathcal{L}:L^{2}\left(\mathbb{S}\right)\longrightarrow L_{f}^{2}\left(\mathbb{A}\right)$,
the binning operator $\mathcal{B}:L_{f}^{2}\left(\mathbb{A}\right)\longrightarrow\mathbb{R}_{f}^{M}$,
and their concatenation into the binned system operator $\mathcal{H}:L^{2}\left(\mathbb{S}\right)\longrightarrow\mathbb{R}_{f}^{M}$.
We have the deomposition in $L_{f}^{2}\left(\mathbb{A}\right)$ of
the function $\mathcal{L}\triangle f$ as f$\mathcal{L}\triangle f=\left(\mathcal{L}\triangle f\right)_{1}+\left(\mathcal{L}\triangle f\right)_{0}$,
where $\mathcal{B}\left(\mathcal{L}\triangle f\right)_{0}=\mathbf{0}$
and $\left(\left(\mathcal{L}\triangle f\right)_{1},\left(\mathcal{L}\triangle f\right)_{0}\right)_{f}=0$. 

As before we find the adjoint of the binning operator, as an operator
between weight Hilbert spaces, via
\begin{equation}
\left(\mathbf{g},\mathcal{B}g'\right)_{f}=\left(\mathbf{g},\mathbf{D}_{f}^{-1}\mathcal{B}g'\right)=\left(\mathcal{B^{\dagger}}\mathbf{D}_{f}^{-1}\mathbf{g},g'\right)=\left(\mathcal{D}_{f}\mathcal{B^{\dagger}}\mathbf{D}_{f}^{-1}\mathbf{g},\mathcal{D}_{f}^{-1}g'\right)=\left(\mathcal{D}_{f}\mathcal{B^{\dagger}}\mathbf{D}_{f}^{-1}\mathbf{g},g'\right)_{f}.
\end{equation}
We then have the pseudoinverse of the binning operator $\mathcal{B}^{+}=\mathcal{D}_{f}\mathcal{B^{\dagger}}\mathbf{D}_{f}^{-1}\left(\mathcal{B}\mathcal{D}_{f}\mathcal{B^{\dagger}}\mathbf{D}_{f}^{-1}\right)^{-1}$,
which simplifies to $\mathcal{B}^{+}=\mathcal{D}_{f}\mathcal{B^{\dagger}}\left(\mathcal{B}\mathcal{D}_{f}\mathcal{B^{\dagger}}\right)^{-1}$.
Computing the operator in parentheses in this last expression leads
to
\begin{equation}
\mathcal{B}\mathcal{D}_{f}\mathcal{B^{\dagger}}\mathbf{g}=\mathcal{B}\mathcal{D}_{f}\sum_{m=1}^{M}g_{m}b_{m}\left(\mathbf{a}\right)=\mathcal{B}\mathcal{L}f\left(\mathbf{a}\right)\sum_{m'=1}^{M}g_{m'}b_{m'}\left(\mathbf{a}\right).
\end{equation}
Examining this equation componentwise then gives us
\begin{equation}
\left(\mathcal{B}\mathcal{D}_{f}\mathcal{B^{\dagger}}\mathbf{g}\right)_{m}=g_{m}\int_{\mathbb{A}}\mathcal{L}f\left(\mathbf{a}\right)b_{m}\left(\mathbf{a}\right)d^{q}a=g_{m}\left(\mathcal{H}f\right)_{m}.
\end{equation}
Therefore we have $\mathcal{B}\mathcal{D}_{f}\mathcal{B^{\dagger}}=\mathbf{D}_{f}$
and the needed pseudoinverse is gien by $\mathcal{B}^{+}=\mathcal{D}_{f}\mathcal{B^{\dagger}}\mathbf{D}_{f}^{-1}$.

Now we have for the first term in the orthogonal decomposition $\left(\mathcal{L}\triangle f\right)_{1}=\mathcal{B}^{+}\mathcal{B}\mathcal{L}\triangle f$.
If we write this equation out explicitly it becomes
\begin{equation}
\left(\mathcal{L}\triangle f\right)_{1}\left(\mathbf{a}\right)=\mathcal{L}f\left(\mathbf{a}\right)\sum_{m=1}^{M}\left[\frac{\left(\mathcal{H}\triangle f\right)_{m}}{\left(\mathcal{H}f\right)_{m}}\right]b_{m}\left(\boldsymbol{a}\right).
\end{equation}
Then the null component of $\mathcal{L}\triangle f$ with respect
to the binning operator in the weighted Hilbert space is $.\left(\mathcal{L}\triangle f\right)_{0}\left(\mathbf{a}\right)=\mathcal{L}\triangle f\left(\mathbf{a}\right)-\left(\mathcal{L}\triangle f\right)_{1}\left(\mathbf{a}\right)$.
Using the orthogonality of the decomposition we have $\left\Vert \mathcal{L}\triangle f\right\Vert _{f}^{2}=\left\Vert \left(\mathcal{L}\triangle f\right)_{1}\right\Vert _{f}^{2}+\left\Vert \left(\mathcal{L}\triangle f\right)_{0}\right\Vert _{f}^{2}$.
The first term in the sum on the right is
\begin{equation}
\left\Vert \left(\mathcal{L}\triangle f\right)_{1}\right\Vert _{f}^{2}=\int_{\mathbb{A}}\left\{ \left[\mathcal{L}\triangle f\left(\mathbf{a}\right)\right]_{1}\right\} ^{2}\left[\mathcal{L}f\left(\mathbf{a}\right)\right]^{-1}d^{q}a
\end{equation}
Using the properties of the bin functions we then have
\begin{equation}
\left\Vert \left(\mathcal{L}\triangle f\right)_{1}\right\Vert _{f}^{2}=\sum_{m=1}^{M}\left[\frac{\left(\mathcal{H}\triangle f\right)_{m}}{\left(\mathcal{H}f\right)_{m}}\right]^{2}\int_{\mathbb{A}}\mathcal{L}f\left(\mathbf{a}\right)b_{m}\left(\boldsymbol{a}\right)d^{q}a
\end{equation}
Thus we have $\left\Vert \left(\mathcal{L}\triangle f\right)_{1}\right\Vert _{f}^{2}=\left(\triangle f,\mathcal{F}_{B}\left(f\right)\triangle f\right)$.

Now we see that the null component $\left(\mathcal{L}\triangle f\right)_{0}$
determines the loss of Fisher information: $\left(\triangle f,\mathcal{F}_{LM}\left(f\right)\triangle f\right)-\left(\triangle f,\mathcal{F}_{B}\left(f\right)\triangle f\right)=\left\Vert \left(\mathcal{L}\triangle f\right)_{0}\right\Vert _{f}^{2}$.
Alternatively we can write
\begin{equation}
\left(\triangle f,\mathcal{F}_{LM}\left(f\right)\triangle f\right)-\left(\triangle f,\mathcal{F}_{B}\left(f\right)\triangle f\right)=\int_{\mathbb{A}}\left\{ \left[\mathcal{L}\triangle f\left(\mathbf{a}\right)\right]_{0}\right\} ^{2}\left[\mathcal{L}f\left(\mathbf{a}\right)\right]^{-1}d^{q}a.
\end{equation}
The two approximate detectabilities are equal only if
\begin{equation}
\mathcal{L}\triangle f\left(\mathbf{a}\right)=\mathcal{L}f\left(\mathbf{a}\right)\sum_{m=1}^{M}\left[\frac{\left(\mathcal{H}\triangle f\right)_{m}}{\left(\mathcal{H}f\right)_{m}}\right]b_{m}\left(\boldsymbol{a}\right)
\end{equation}
This condition implies that for almost all perturbation functions
$\triangle f\left(\mathbf{a}\right)$ the list-mode approximate detectability
will be greater than the binned approximate detectability. 

Note that the condition for no loss of information due to binning
can also be written as 
\begin{equation}
\left(\mathcal{L}\triangle f\right)_{0}\left(\mathbf{a}\right)=\mathcal{L}f\left(\mathbf{a}\right)\sum_{m=1}^{M}\left[\frac{\mathcal{L}\triangle f\left(\mathbf{a}\right)}{\mathcal{L}f\left(\mathbf{a}\right)}-\frac{\left(\mathcal{H}\triangle f\right)_{m}}{\left(\mathcal{H}f\right)_{m}}\right]b_{m}\left(\boldsymbol{a}\right)=0.
\end{equation}
This then gives us
\begin{equation}
\left(\triangle f,\mathcal{F}_{LM}\left(f\right)\triangle f\right)-\left(\triangle f,\mathcal{F}_{B}\left(f\right)\triangle f\right)=\sum_{m=1}^{M}\int_{\mathbb{A}}\left[\frac{\mathcal{L}\triangle f\left(\mathbf{a}\right)}{\mathcal{L}f\left(\mathbf{a}\right)}-\frac{\left(\mathcal{H}\triangle f\right)_{m}}{\left(\mathcal{H}f\right)_{m}}\right]^{2}\mathcal{L}f\left(\mathbf{a}\right)b_{m}\left(\boldsymbol{a}\right)d^{q}a.
\end{equation}
Thus, as in the case described above for a finite dimensional parameter,
each bin contributes to the loss of the detectability of a change
in the object function according to three factors. The first factor
is again the deviation of the quantity in curly brackets from zero
within that bin. The second factor is the value of the function $\mathcal{L}f\left(\mathbf{a}\right)$
within the bin. The third factor is the size of the bin itself. The
efficiency of any particular choice of bins in preserving Fisher information
depends on the actual object function $f$ as well as bin size.

Finally, note that, as in the previous section, this last equality
can be proved directly. Again, the path follwed in this derivation
shows that the loss of Fisher information about the object function
due to binning comes from the null space of $\mathcal{B}:L_{f}^{2}\left(\mathbb{A}\right)\longrightarrow\mathbb{R}_{f}^{M}$
as an operator between weighted Hilbert spaces.

\section{Example}

For this example, consider the attribute space to be a symmetric interval
on the real line: $\mathbb{A}=\left[-L/2,L/2\right]$. The object
functions will be square integrable functions of a real variable and
the list-mode system operator is convolution with a pint spread function
(PSF): $\mathcal{L}f\left(x\right)=p\ast f\left(x\right)$. We assume
that the point spread function is band limited to the band $\left[-B/2,B/2\right]$. 

Now let $M$ and $\triangle x$ be such that $L=M\triangle x$ and
define the regularly spaced points in $\mathbb{A}$ via
\begin{equation}
x_{m}=-\frac{L}{2}+\left(m-\frac{1}{2}\right)\triangle x
\end{equation}
and the bin functions as
\begin{equation}
b_{m}\left(x\right)=\mathrm{rect}\left(\frac{x-x_{m}}{\triangle x}\right).
\end{equation}
We now have the binning operator described by
\begin{equation}
\left(\mathcal{B}g\right)_{m}=\int g\left(x\right)\mathrm{rect}\left(\frac{x-x_{m}}{\triangle x}\right)dx=\int_{x_{m}-\frac{\triangle x}{2}}^{x_{m}+\frac{\triangle x}{2}}g\left(x\right)dx.
\end{equation}

The condition for no loss in the approximate detectability by binning
is given by
\begin{equation}
p\ast\triangle f\left(x\right)=p\ast f\left(x\right)\sum_{m=1}^{M}\left[\frac{\left(\mathcal{H}\triangle f\right)_{m}}{\left(\mathcal{H}f\right)_{m}}\right]\mathrm{rect}\left(\frac{x-x_{m}}{\triangle x}\right)
\end{equation}
This condition is impossible to satisfy since the function on the
left is band-limited and the function on the right, in general, is
not. Thus, even with Nyquist sampling, when $B\triangle x=1$, there
is a loss in the detectability of a small change in the object function
when we bin the list-mode data. The actual loss of Fisher information
for a small change in the object function is given by
\begin{equation}
\left(\triangle f,\mathcal{F}_{LM}\left(f\right)\triangle f\right)-\left(\triangle f,\mathcal{F}_{B}\left(f\right)\triangle f\right)=\sum_{m=1}^{M}\int_{x_{m}-\frac{\triangle x}{2}}^{x_{m}+\frac{\triangle x}{2}}\left\{ \frac{p\ast\triangle f\left(x\right)}{p\ast f\left(x\right)}-\left[\frac{\left(\mathcal{H}\triangle f\right)_{m}}{\left(\mathcal{H}f\right)_{m}}\right]\right\} ^{2}p\ast f\left(x\right)dx.
\end{equation}
In general, loss of Fisher information is mitigated if $B$ is decreased
since this will mean that $p\ast\triangle f\left(x\right)$ and $p\ast f\left(x\right)$
are smoother functions, and hence there will be a decrease the quantities
in the curly brackets. 

There is at least one circumstance in this example where there is
no loss of Fisher information from binning the list-mode data. If
$\triangle f\left(x\right)=\alpha f\left(x\right)$ for some constant
$\alpha$, then $\left(\triangle f,\mathcal{F}_{LM}\left(f\right)\triangle f\right)-\left(\triangle f,\mathcal{F}_{B}\left(f\right)\triangle f\right)=0.$
his is true even if $M=1$ and $\triangle x=L$. In other words, to
detect a simple change in amplitude of the object function we might
as well use one bin covering all of $\mathbb{A}$. There may also
be other special situations where binning does not create a loss of
Fisher information, but for generic functions $f\left(x\right)$and
$\triangle f\left(x\right)$ there will always be a loss.

\section{Conclusion}

We have shown that there is almost always a loss of Fisher information
for any estimatioion task when list-mode data is binned. This loss
of information is due to the null space of the binning operator when
it is viewed as an operator between certain parameter dependent weighted
Hilbert spaces. The magnitude of the loss can be quantified by finding
the null component, with respect to the binning operator, of a directional
derivative of the conditional PDF as an element of one of the weighted
Hilbert spaces. We found that the information loss depends on the
smoothness of the mean data function for the list-mode data, the actual
object being imaged, and the the binning scheme in relation to the
other two factors. We have shown that these conclutions apply even
when the estimation problem is an object reconstruction problem, where
the finite dimensional parameter vector is replaced with a function
in an infinite dimensional Hilbert space. 

As a final note the difference $\triangle\boldsymbol{\theta}^{\dagger}\mathbf{F}_{LM}\left(\boldsymbol{\theta}\right)\triangle\boldsymbol{\theta}-\triangle\boldsymbol{\theta}^{\dagger}\mathbf{F}_{B}\left(\boldsymbol{\theta}\right)\triangle\boldsymbol{\theta}$
can be written as
\begin{equation}
\sum_{m=1}^{M}\int_{\mathbb{A}}\triangle\boldsymbol{\theta}^{\dagger}\left\{ \frac{\left[\nabla_{\boldsymbol{\theta}}\bar{g}\left(\mathbf{a}|\boldsymbol{\theta}\right)\right]\left[\nabla_{\boldsymbol{\theta}}\bar{g}\left(\mathbf{a}|\boldsymbol{\theta}\right)\right]^{\dagger}}{\bar{g}\left(\mathbf{a}|\boldsymbol{\theta}\right)}-\frac{\left[\nabla_{\boldsymbol{\theta}}\bar{g}_{m}\left(\boldsymbol{\theta}\right)\right]\left[\nabla_{\boldsymbol{\theta}}\bar{g}_{m}\left(\boldsymbol{\theta}\right)\right]^{\dagger}}{\bar{g}_{m}\left(\boldsymbol{\theta}\right)}\right\} \triangle\boldsymbol{\theta}^{\dagger}\bar{g}\left(\mathbf{a}|\boldsymbol{\theta}\right)b_{m}\left(\boldsymbol{a}\right)d^{q}a.
\end{equation}
Therefore we have an expression for the difference $\mathbf{F}_{LM}\left(\boldsymbol{\theta}\right)-\mathbf{F}_{B}\left(\boldsymbol{\theta}\right)$
of FIMs:
\begin{equation}
\sum_{m=1}^{M}\int_{\mathbb{A}}\left\{ \frac{\left[\nabla_{\boldsymbol{\theta}}\bar{g}\left(\mathbf{a}|\boldsymbol{\theta}\right)\right]\left[\nabla_{\boldsymbol{\theta}}\bar{g}\left(\mathbf{a}|\boldsymbol{\theta}\right)\right]^{\dagger}}{\bar{g}\left(\mathbf{a}|\boldsymbol{\theta}\right)}-\frac{\left[\nabla_{\boldsymbol{\theta}}\bar{g}_{m}\left(\boldsymbol{\theta}\right)\right]\left[\nabla_{\boldsymbol{\theta}}\bar{g}_{m}\left(\boldsymbol{\theta}\right)\right]^{\dagger}}{\bar{g}_{m}\left(\boldsymbol{\theta}\right)}\right\} \bar{g}\left(\mathbf{a}|\boldsymbol{\theta}\right)b_{m}\left(\boldsymbol{a}\right)d^{q}a.
\end{equation}
Now if we have a nominal value for $\boldsymbol{\theta}$, but there
is some uncertainty in this value, then this is equivalent to making
$\triangle\boldsymbol{\theta}$ a random vector with zero mean. If
the covariance matrix for this vector is $\mathbf{K}_{\boldsymbol{\theta}}$
then the average value for $\triangle\boldsymbol{\theta}^{\dagger}\mathbf{F}_{LM}\left(\boldsymbol{\theta}\right)\triangle\boldsymbol{\theta}-\triangle\boldsymbol{\theta}^{\dagger}\mathbf{F}_{B}\left(\boldsymbol{\theta}\right)\triangle\boldsymbol{\theta}$
is $\mathrm{tr}\left\{ \mathbf{K}_{\theta}\left[\mathbf{F}_{LM}\left(\boldsymbol{\theta}\right)-\mathbf{F}_{B}\left(\boldsymbol{\theta}\right)\right]\right\} $.
This may be a useful quantification of the average loss of Fisher
information due to binning in this situation. When $\mathbf{K}_{\boldsymbol{\theta}}=\sigma^{2}\mathbf{I}$
we end up with
\[
\mathrm{tr}\left\{ \mathbf{K}_{\theta}\left[\mathbf{F}_{LM}\left(\boldsymbol{\theta}\right)-\mathbf{F}_{B}\left(\boldsymbol{\theta}\right)\right]\right\} =\sigma^{2}\sum_{m=1}^{M}\int_{\mathbb{A}}\left[\frac{\left\Vert \nabla_{\boldsymbol{\theta}}\bar{g}\left(\mathbf{a}|\boldsymbol{\theta}\right)\right\Vert ^{2}}{\bar{g}\left(\mathbf{a}|\boldsymbol{\theta}\right)}-\frac{\left\Vert \nabla_{\boldsymbol{\theta}}\bar{g}_{m}\left(\boldsymbol{\theta}\right)\right\Vert ^{2}}{\bar{g}_{m}\left(\boldsymbol{\theta}\right)}\right]\bar{g}\left(\mathbf{a}|\boldsymbol{\theta}\right)b_{m}\left(\boldsymbol{a}\right)d^{q}a.
\]
This is a relatively compact expression that can be easily evaluated
in many cases.

\end{document}